\documentclass[runningheads]{llncs}
\usepackage{latexsym,amsmath}
\usepackage{amssymb}

\begin{document}


\title{Linear Intransitive Temporal Logic of Knowledge $LTK_r$, Decision Algorithms, Inference Rules}


\author{ Alexandra Lukyanchuk$^2$, Vladimir Rybakov$^1$}
\authorrunning{A. Lukyanchuk, V.Rybakov}
\titlerunning{Linear Intransitive Temporal Logic of Knowledge and Time $LTK_r$ }

\institute{(1)School of Computing, Mathematics and DT,
  Manchester Metropolitan University,
 John Dalton Building, Chester Street, Manchester, M1 5GD, U.K.  \\
 (2)Institute of Mathematics and Information Technologies, Siberian Federal University, Krasnoyarsk, Russia
 \email{}
 }


\mainmatter

\maketitle

\centerline{Email: V.Rybakov@mmu.ac.uk}

\newcommand{\ga}{\alpha}
\newcommand{\gb}{\beta}
\newcommand{\grg}{\gamma}
\newcommand{\gd}{\delta}
\newcommand{\gl}{\lambda}
\newcommand{\cff}{{\cal F}or}
\newcommand{\ca}{{\cal A}}
\newcommand{\cb}{{\cal B}}
\newcommand{\cc}{{\cal C}}
\newcommand{\cm}{{\cal M}}
\newcommand{\cmm}{{\cal M}}
\newcommand{\cbb}{{\cal B}}
\newcommand{\ccrr}{{\cal R}}
\newcommand{\cf}{{\cal F}}
\newcommand{\cy}{{\cal Y}}
\newcommand{\cxx}{{\cal X}}
\newcommand{\cdd}{{\cal D}}
\newcommand{\cww}{{\cal W}}
\newcommand{\czz}{{\cal Z}}
\newcommand{\cll}{{\cal L}}
\newcommand{\cw}{{\cal W}}
\newcommand{\ckk}{{\cal K}}
\newcommand{\ppp}{{\varphi}}

\newcommand{\ii}[0]
{\rightarrow}

\newcommand{\ri}[0]
{\mbox{$\Rightarrow$}}

\newcommand{\lri}[0]
{\mbox{$\Leftrightarrow$}}

\newcommand{\lr}[0]
{\mbox{$\Longleftrightarrow$}}

\newcommand{\ci}[1]{\cite{#1}}

\newcommand{\pr}{{\sl Proof}}

\newcommand{\vv}[0]{
\unitlength=1mm
\linethickness{0.5pt}
\protect{
\begin{picture}(4.40,4.00)
\put(1.2,-0.4){\line(0,1){3.1}}
\put(2.1,-0.4){\line(0,1){3.1}}
\put(2.1,1.1){\line(1,0){2.0}}
\end{picture}\hspace*{0.3mm}}}

\newcommand{\nv}[0]{
\unitlength=1mm
\linethickness{0.5pt}
\protect{
\begin{picture}(4.40,4.00)
\put(1.2,-0.4){\line(0,1){3.1}}
\put(2.1,-0.4){\line(0,1){3.1}}
\put(2.1,1.1){\line(1,0){2.0}}
\protect{
\put(0.3,-0.7){\line(1,1){3.6}}}
\end{picture}\hspace*{0.3mm}
}
}

\newcommand{\nvv}[0]{
\unitlength=1mm
\linethickness{0.5pt}
\protect{
\begin{picture}(4.40,4.00)
\put(2.1,-0.4){\line(0,1){3.1}}
\put(2.1,1.2){\line(1,0){2.0}}
\protect{
\put(0.6,-0.5){\line(1,1){3.6}}}
\end{picture}\hspace*{0.3mm}
}
}

\newcommand{\dd}[0]{
\rule{1.5mm}{1.5mm}}

\newcommand{\llll}{{\cal LT \hspace*{-0.05cm}L}}

\newcommand{\nn}{{\bf N}}

\newcommand{\pp}{{{\bf N^{-1}}}}

\newcommand{\uu}{{{\bf U}}}

\newcommand{\suu}{{{\bf S}}}

\newcommand{\bb}{{{\bf B}}}

\newcommand{\nnn}{{\cal N}}
\newcommand{\sss}{{{\bf S}}}
\newcommand{\zzz}{{\cal LT \hspace*{-0.05cm}L}_K(Z)}
\newcommand{\zz}{{{\cal Z}_C}}

\date{}

\bigskip

\begin{abstract} Our paper investigates the linear logic of knowledge and time $LTK_r$ with reflexive intransitive time relation. The logic is defined semantically, -- as the set of formulas true at special frames 
with intransitive and reflexive time binary relation. The $LTK_r$-frames are linear chains of clusters connected by a reflexive intransitive relation $R_T$. Elements inside a cluster are connected by several equivalence relations imitating the knowledge of different agents. We study the 
decidability problem for formulas and inference rules. Decidability for formulas follows from decidability w.r.t. admissible inference rules.To study admissibility, we 
introduce some
special constructive Kripke models suitable to describe admissibility of inference rules. 
With a special technique of definable valuations we  
find an algorithm determining admissible inference rules in $LTK_r$.. That is, we show that the logic $LTK_r$ 
 is decidable and decidable with respect to admissibility of inference rules.
     \end{abstract}

\bigskip


\hspace*{1.8cm} Algorithms, $n$-characterizing models, Admissible rules

\bigskip

\section{Introduction}

Interest to study of inference rules in non-standard epistemic logics appeared quite a while ago primarily from proof theory and its applications to
 computer sciences (CS). Research of artificial intelligence (AI) requires language adapted to description of various dynamic systems. The language of multi-modal logic, combining temporal and  knowledge modalities, perfectly cope with this task (\cite{fag1}, \cite{tom1}). Multi-modal logics generated by adjoining
operators representing time and knowledge to the classical propositional calculus ${\bf CPC}$ are very effective for modeling reasoning (in particular, where agents, who possess a certain knowledge, are operating in the processes of computation in a flow of time (\cite{fag1},  \cite{hal1}). But initially, the facts and statements are described by formulas, and just formulas themselves are not always able to express the changing conditions and prerequisites in a very effective manner. We often want to know want will follow from given statements (assumptions), and for this we need to know what are correct logical consequence for given assumptions. So we can extend the language of logic by considering conditional statements of the form $A_1, \dots, A_n / B$, which have the meaning {\em if all $A_1, \dots, A_n$ hold (are true), then $B$ also holds (is true)}. 

We will study logical consequence in terms of admissible and valid inference rules.
The notion of admissible inference rules was introduced by Lorenzen in 1955 (\cite{lor}). Admissible rules of a logic are those ones under which the set of theorems of this logic is closed. 
It was observed, that we can expand an axiomatic system by adding admissible, though not derivable, inference rules e.g. for Heyting intuitionistic logic ${\bf IPC}$  (\cite{har}). Hence the question of finding algorithm to determine admissible rules in non-classical logics was set up. Originally H. Friedman addressed this question  to the intuitionistic logic ${\bf IPC}$ (\cite{frid}) itself. This problem  has been solved by V. Rybakov (\cite{ryb8,ryb7}). Then the admissibility question was studied for many other non-standard logics (\cite{ryb6,gol1,cal1}). 

S. Ghilardi has found algorithm recognizing admissibility  using the concept of unification(\cite{gil1}). Later Vladimir Rybakov investigated the unification problem and its connection with the of decidability w.r.t inference rules in class of many popular logics (\cite{ryb1,ryb2,ryb3,ryb4,ryb5}).

 In this paper, we extend the investigation of this area to a linear temporal multi-modal logic $LTK_r$ with linear intransitive and reflexive time and agent's knowledge studied in \cite{luk1,luk2}. 
We consider the time as a linear and discrete sequence of states. Each state consists of a set of information points connected by modal relations $R_i$. In other words, $R_i$ says which information points are effectively available for the agent $i$: it species the piece of information that the agent may access at given moment. Agents operating synchronously and each agent knows what time it is and distinguishes present state from the next one.
  The prime question we are dealing with in this article is the decidability of $LTK_r$. We reduce the decidability problem for $LTK_r$ to validness verification for inference rules $r$ in reduced normal form in special Kripke models, whose size is computable and bounded by the size of $r$. Hence, we prove that $LTK_r$ is decidable w.r.t. admissible inference rules and w.r.t. true formulas (theorems).

\section{Necessary Preliminary Information}

First we recall some necessary basic definitions, notation and known results used in this article.

The language ${\mathcal L}^{LTK}$ consists of a countable set of propositional letters $P:=\{p_1, \dots, p_n, \dots \}$, the standard boolean operations and the set of modal operations $\{ \Box_T, \Box_{\sim}, \Box_i$ $(i \in I) \}$. {\it Well formed formulae} (wff's) are defined in the standard way, in particular, if $A$ is a wff, than $\Box_T A, \Box_{\sim} A, \Box_i A (i \in I)$ are wff's. We denote by $Fma({\mathcal L}^{LTK})$  the set of all the wff's of ${\mathcal L}^{LTK}$ (in the sequel, in saying -  {\it formula} we always refer to a formula from $Fma({\mathcal L}^{LTK})$). The intended meaning of the modal operations is: (a) $\Box_T A$ for logic $LTK_r$ means that the formula $A$ true in the current state and will be true in the next state. (b) $\Box_{\sim} A$ means that $A$ is known everywhere in the present time-cluster (i.e. $A$ is part of the environmental knowledge); (c) $ \Box_i A (i\in I)$ stands for the agent $i$ (operating in the system) knows $A$ in the current state. Semantics for the language ${\mathcal L}^{LTK}$ is based on a linear and discrete flow of time, associating a time point with any natural number $n$.

An {\it $LTK_r$-frame} is a multi-modal frame $\cf =\langle W_{\cf},R_T, R_{\sim}, R_1, \dots, R_k \rangle$, 

 \ \ \ where:

(a) $W_{\cf}$ is the disjoint union of certain nonempty sets $C_n$: $W_{\cf}:=  \bigcup\limits_{n \in J} C_n$ 

\ \ \ where 
$J =[0, L]$ and $L \in N$ or $J=N$.

\medskip

(b) $R_T$ is the linear, reflexive and intransitive relation on $W_{\cf}$ such that:

\[\forall w \forall z \in W_{\cf} (w R_T z \Leftrightarrow [\exists n \in J ((w \in C_n) \& \]

\[(z \in C_n))] \vee  [\exists n+1 \in J((w \in C_n) \& (z \in C_{n+1}))])\]

(c) $R_{\sim}$ is a universal relation on any $C_n \in W_{\cf}$:

 \ \ \ $\forall w \forall z \in W_{\cf} (w R_{\sim} z \Leftrightarrow \exists n \in J ((w \in C_n) \& (z \in C_n)));$

\medskip

(d)  $ \forall i \in I, R_i$ is some equivalence relation on $C_n$.

\medskip

Let ${\sf LTK_r}$ be the class of all $LTK_r-$frames.

\medskip

Such frames simulate the situation in which agents, having a certain knowledge background at a given moment, are operating in the linear flow of time. Each time-cluster (i.e. an $R_T-$cluster) $C_n$ consists of a set of information points that are available at the moment $n$. The relation $R_T$ is the connection of such information points by the flow of time.
 That is, given two information points $w$ and $z$, the expression $w R_T z$ means either that $w$ and $z$ are both available at a moment $n$, or that $z$ will be available in the moment $n+1$ with respect to $w$. Since the relation $R_{\sim}$ connects all the information-points available at the same moment, it is intended to represent a sort of environmental knowledge, that is, the whole information potentially available for the agent at a given time. The relation $R_i$ says which information points are effectively available for the agent $i$ at any given moment.
  
Moreover, relations on $LTK_r$-frame possess the following properties:

PM.1: $vR_{\sim}z \Longrightarrow$ $(vR_Tz$ $\&$ $zR_Tv)$

PM.2: $vR_iz \Longrightarrow$ $vR_{\sim}z $

PM.3: $(vR_Tz$ $\&$ $zR_Tv)\Longrightarrow$ $vR_{\sim}z$.

In particular, the coincidence of the $R_T$- and $R_{\sim}$-clusters of the linear chain is assumed [15].

Given a model $\cm = \langle \cf, V \rangle$, where $\cf$ is an $LTK_r-$ frame, the valuation $V$ can be extended in the standard way from the set $P$ of propositional letters to all well formed formulae constructed from $P$.

\begin{definition} Computational rules for logical operations:

\begin{itemize}

\item $ \forall p\in P \ ( {\mathcal F}, w) \models_V p\  \ \ \lri  \ \ \ w \in V(p);$

\item $ ({\mathcal F}, w) \models_V \Box_T A \ \ \ \lri \ \ \ $
$\forall z \in W_{{\mathcal F}} (w R_T z \Rightarrow ({\mathcal F}, z) \models_V A);$

\item $ ({\mathcal F}, w) \models_V \Box_{\sim} A \ \ \  \lri \ \ \ $
$ \forall z \in W_{{\mathcal F}} (w R_{\sim} z \Rightarrow ( {\mathcal F}, z) \models_V A);$

\item  $ \forall i \in I, ( {\mathcal F}, w) \models_V \Box_i A \ \ \ \lri \ \ \ \ $
$\forall z \in W_{{\mathcal F}} (w R_i z \Rightarrow ( {\mathcal F}, z) \models_V A).$

\end{itemize}

\end{definition}

Logic $LTK_r$ is the set of all $LTK_r-$valid formulae:

$$LTK_r:=\{ A \in Fma({\mathcal L}^{LTK_r}) | \forall \cf \in {\sf LTK_r }(\cf \models A )\}.$$

 If $A$ belongs to $LTK_r$, then $A$ is said to be a theorem of $LTK_r$.

\begin{definition} {\it Time degree $td(A)$ of a formula $A$}  is defined as follows: $td(p)=td(T)=td(\perp)=0$; $td(\neg \ga)=td(\ga)$; $td(\ga \to \gb)=td(\ga \vee \gb)=td(\ga \wedge \gb)=max(td(\ga),td(\gb))$;  $td(\Box_{\sim} \ga)=td(\Box_i \ga)=td(\ga)$; $td(\Box_T \ga)=td(\ga)+1$.

\end{definition}

\begin{definition} Given a logic $L$, a model $Ch_L(n) := \langle Ch(n) ,V \rangle$
is  said to be {\it an $n$-characterizing model for $L$} iff: $(a) Dom (V) :=\{p_1, \dots, p_n\}$ $(b)$ for any formula $A$ built up from $Dom(V)$, $Ch(n)  \models_V A \Leftrightarrow A \in L$.

\end{definition} 

\begin{definition} Given a model $\langle Ch(n) ,V \rangle$, a world $w \in W_{Ch(n)}$  is{ \it definable } iff there is a formula $\gb(w)$ such that $\forall z \in Ch(n) (Ch(n),z) \models_V \gb(w) \Leftrightarrow w=z)$.

\end{definition} 

{\it A consecution (an inference rule)} $r$ is an expression
 
 $$r:=\frac{\varphi_1(x_1,\dots,x_m), \dots, \varphi_n(x_1,\dots,x_m)}{\phi(x_1,\dots,x_m)},$$
where $\varphi_1(x_1,\dots,x_m), \dots, \varphi_n(x_1,\dots,x_m)$ and $\phi(x_1,\dots,x_m)$ are wff build up from the letters $x_1, \dots, x_m$.  Expression $Pr(r)$ is an abbreviation for the premises of $r$, and $Con(r)$ for the conclusion of $r$.  

An inference rule $r:= \varphi_1(x_1,\dots,x_m), \dots, \varphi_n(x_1,\dots,x_m) / \phi(x_1,\dots,x_m)$ is  {\it admissible} for a logic $L$ ( $r \in  Ad(L)$) iff for each substitution $\Sigma$, if  $\Sigma(\varphi_i) \in L$ for each $i$, then $\Sigma(\phi) \in L$.

A rule $r$ is in the {\it reduced normal form} if $r:=\epsilon_r/x_0$, where

$$\epsilon_r:=\bigvee_{1\le j\le s} \theta_j; \theta_j :=(\bigwedge_{1\le i\le m} [x_i^{d(j,i,1)} \wedge$$

$$(\Diamond_T x_i)^{d(j,i,2)} \wedge (\Diamond_{\sim} x_i)^{d(j,i,3)} \wedge \bigwedge_{1\le l\le k}(\Diamond_{l} x_i)^{d(j,i,l,4)}]),$$

$d(j,i,z), d(j,i,l,z) \in \{0, 1\}$ and for any formula $\alpha$ above, $\alpha^0:=\alpha$, $\alpha^1 := \neg\alpha$.

\medskip

Given a rule $r_{nf}$ in the reduced normal form, $r_{nf}$ is said to be a {\it normal reduced form for a rule $r$} iff, for any frame $\cf$, $\cf \models r \Leftrightarrow \cf \models r_{nf}$. 

Using Corollary 3.1.13 and Corollary 3.1.15 from \cite{ryb6}, we obtain:

\begin{theorem}

There exist an algorithm running in (single) exponential time, which, for any given rule $r$ in the language of logic $LTK_r$, constructs
its normal reduced form $r_{nf}$. Moreover, $r$ is semantically equivalent to $r_{nf}$ in $LTK_r$.
 
\end{theorem}

\section{Construction of $Ch_{LTK_r}(n)$ }

In this section we will construct special $n$-characterizing models for the logic $LTK_r$ in case of intransitive time relation. This construction based on the techniques presented by V.V. Rybakov in \cite{ryb6}.

 {\it Step 1.}

Let $F$ be a set of finite $LTK_r$-frames such that, for any frame $\cf \in F, \forall w \forall z \in W_{\cf} (w R_T z$ $\&$ $z R_Tw)$. Let ${\mathcal C}(F)_n$ be a set of all different, non-isomorphic models $C := \langle \cf, V \rangle$, where

\begin{enumerate}

 \item $\cf \in F;$

 \item $Dom(V) = \{p_1, \dots, p_n \};$

     \end{enumerate}

Let $S_1(Ch_{(LTK_r)}(n)) := \bigsqcup\limits_{{\mathcal C} (F)} C_n$, the first slice of $Ch_{LTK_r}(n)$ contains a finite number of $R_T$-clusters with valuation of variables $p_1, \dots, p_n$ s.t.$\forall C_i, C_j \in {\mathcal C}(F),C_i$ is not isomorphic to $C_j$.

  {\it Step 2.}

To each $C$ from $S_1(Ch_{(LTK_r)}(n))$ we adjoin non-isomorphic to $C$ models $C_j$ from ${\mathcal C}(F)_n$ assuming $C_j$ to be immediate $R_T$-predecessor of $C$. The resulting model is defined as $S_{\le 2}(Ch_{(LTK_r)}(n))$.

 {\it Step 3.}
 
 To each $C$ from $S_2(Ch_{(LTK_r)}(n))$ we adjoin all models $C_j$ from ${\mathcal C}(F)_n$ as immediate $R_T$-predecessor of $C$. The resulting model is defined as $S_{\le 3}(Ch_{(LTK_r)}(n))$.

 {\it Step 4.}
 
 Suppose, we have already constructed the model $S_{\le i}(Ch_{(LTK_r)}(n))$ for $i \ge 2$ such that its frame is is an $LTK_r$-frame.
 
 To construct $S_{\le i+1}(Ch_{(LTK_r)}(n))$ we add all models from ${\mathcal C}(F)_n$ to each $R_T$-cluster from $S_i(Ch_{(LTK_r)}(n))$ as its immediate $R_T$-predecessors.
 
The resulting models of such extension is the model \\ 

$$Ch_{LTK}(n) :=\langle W_{Ch_{(LTK_r)}}, R_T, R_{\sim}, R_1, \dots, R_k, V \rangle := $$

$$\bigcup\limits_{i \in N} S_{\le i}(Ch_{(LTK_r)}(n)).$$

We will denote the base set of $Ch_{LTK}(n)$ as $Ch(n)$.

    \begin{lemma}
    
The model $Ch_{LTK_r}(n) = \langle Ch(n), V \rangle$ is $n$-characterizing for $LTK_r$.

  \end{lemma}

  \begin{lemma}
     
   For any $n$-characterizing model $Ch_{LTK_r}(n)$, each world $w$ from $W_{Ch(n)}$ is not definable.

  \end{lemma}

\section{Decidability with respect to admissible inference rules}

First we introduce a special kind of $LTK_r$-frames, which plays a leading role in the description of the main result.

\medskip

Let $\cf_p$, $\cf_S$ and $\cf_i$ be $LTK_r$-frames with the following structures:

\medskip

(a) The frame $\cf_P = \left\langle W_{\cf_P}, R^P_T, R^P_{\sim}, R^P_1, \dots, R^P_k\right\rangle$ is an $LTK_r$-frame such that its base set $W_{\cf_P}$ consists only one world denoted by $@$, $W_{\cf_P} := \{@\}$. 

\medskip
 
 (b) Let $\cf_S = \left\langle W_{\cf_S}, R^S_T, R^S_{\sim}, R^S_1, \dots, R^S_k\right\rangle$ be a finite $LTK_r$-frame, where $W_{\cf_S} = \{\bigcup_{i=0}^d C_i\}$ and $C_0R_TC_1R_T\dots R_TC_d$.

\medskip
 
 (c) The frame $\cf_i = \left\langle W_{\cf_i}, R^i_T, R^i_{\sim}, R^i_1, \dots, R^i_k\right\rangle$ is a finite $LTK_r$-frame, and each $R_T^i$-cluster of $\cf_i$ consists of only one world. Namely, $W_{\cf_i}= \{w_1^i, \dots, w^i_{J_i} \}$,  and  $w_1^i R_T w_2^i R_T \dots R_T w_{J_i}^i$.

\medskip 
 
\begin{definition} An {\it $SP$-frame} is a tuple $\cf_{SP} = \left\langle W_{SP}, R_T, R_{\sim}, R_1,\dots, R_k\right\rangle$ where
\medskip

 1) $W_{SP} = W_{\cf_P} \cup W_{\cf_S} \cup \bigcup_{i=0}^d W_{\cf_i}$;
 
 2) $R_T = R^P_T \cup R^S_T \cup \bigcup_{i=0}^d R^i_T \cup \{\left\langle z, @\right\rangle| z \in C_d\} \cup \bigcup_{i=0}^d \{ \left\langle w_{J_i}^i, z\right\rangle | w_{J_i}^i $ 

$\mbox{is }  R_T\mbox{-maximal }$ $\mbox{world of } \cf_i, z \in C_i \subseteq \cf_S\}$;
 
 3) $R_{\sim} = R_{\sim}^{P} \cup R_{\sim}^{S} \cup \bigcup_{i=0}^d R_{\sim}^{i}$; 
 
 4) $R_j = R_j^{P} \cup R_j^{S} \cup \bigcup_{i=0}^d R_j^{i}$ $(1 \le j \le k)$.
 
\end{definition}

\begin{theorem} An inference rule $r_{nf}$ in the reduced normal form is not admissible in $LTK_r$ if and only if there is a finite $SP$-frame $\cf_{SP}$, whose size is computable in the size of $r_{nf}$, and a valuation $V$ for variables from $r_{nf}$ in $\cf_{SP}$, such that 
  
  1) $\cf_{SP} \not\models_V  Con(r_{nf})$; 
  
  2) $\cf_{SP} \models_V Pr(r_{nf})$; 
  
  3)There is $\theta_a \in  Pr(r_{nf})$, where
  
   $$(\cf_{SP},w^i_1) \models_V \theta_a,(\cf_{SP},w^i_2) \models_V \theta_a,(\cf_{SP}, @) \models_V \theta_a,$$ 
  
   for $(0 \le i \le d)$;
   
  4)$\forall z, w \in C_d $ $\&$ $(z\neq w):$ $(\cf_{SP},z) \models_V \theta_k, (\cf_{SP}, w) \models_V  \theta_m$ and $\theta_k \neq \theta_m$ 
    
  5) $R_T$-cluster $C_d$ is not isomorphic to the world  $@$.
  
\end{theorem}

Based on this result we immediately derive

\begin{theorem}

The logic $LTK_r$ is decidable w.r.t. admissible rules (and consequently w.r.t. theorems).

\end{theorem}


\end{document}